\documentstyle[epsf,prl,aps,twocolumn,floats]{revtex}
\begin{document} 
\title{Improved surrogate data for nonlinearity tests}
\date{Phys.\ Rev.\ Lett.\ {\bf 77}, 635 (1996)}
\author{Thomas Schreiber  \and Andreas Schmitz\\
      Physics Department, University of Wuppertal,\\
      D--42097 Wuppertal, Germany}
\maketitle
\begin{abstract} 
Current tests for nonlinearity compare a time series to the null hypothesis of
a Gaussian linear stochastic process. For this restricted null assumption,
random surrogates can be constructed which are constrained by the linear
properties of the data.  We propose a more general null hypothesis allowing for
nonlinear rescalings of a Gaussian linear process. We show that such rescalings
cannot be accounted for by a simple amplitude adjustment of the surrogates
which leads to spurious detection of nonlinearity. An iterative algorithm is
proposed to make appropriate surrogates which have the same autocorrelations as
the data {\em and} the same probability distribution.
\end{abstract}

The paradigm of deterministic chaos has become a very attractive concept for
the study of the irregular time evolution of experimental or natural phenomena.
Nonlinear methods have indeed been successfully applied to laboratory data from
many different systems~\cite{ecc}. However, soon after the first signatures of
low dimensional chaos had been reported for field data~\cite{nico}, it turned
out that nonlinear algorithms can mistake linear correlations, in particular
those of the power law type, for determinism.\cite{theilerremarks} This has
lead on the one hand to more critical applications of algorithms like the
correlation dimension.\cite{ks} On the other hand, significance tests have been
proposed which allow for the detection of nonlinearity even when for example a
clear scaling region is lacking in the correlation integral.\cite{bdsetc} The
idea is to test results against the null hypothesis of a specific class of
linear random processes.

One of the most popular of such tests is the method of ``surrogate data'',
\cite{surro} which can be used with any nonlinear statistic that characterizes
a time series by a single number. The value of the nonlinear discriminating
statistic is computed on the measured data and compared to its empirical
distribution on a collection of Monte Carlo realizations of the null
hypothesis. Usually, the null assumption we want to make is not a very specific
one, like a certain particular autoregressive (AR) process. We would rather
like to be able to test general assumptions, for example that the data is
described by {\em some} Gaussian linear random process. Thus we will not try to
find a specific faithful model of the data; we will rather design the Monte
Carlo realizations to have the same linear properties as the data. The authors
of~\cite{montecarlo} call this a ``constrained realization" approach.

In particular, the null hypothesis of autocorrelated Gaussian linear noise can
be tested with surrogates which are by construction Gaussian random numbers but
have the same autocorrelations as the signal. Due to the Wiener--Khinchin
theorem, this is the case if their power spectra coincide. One can multiply the
discrete Fourier transform of the data by random phases and then perform the
inverse transform (phase randomized surrogates).  Equivalently, one can create
Gaussian independent random numbers, take their Fourier transform, replace
those amplitudes with the amplitudes of the Fourier transform of the original
data, and then invert the Fourier transform.  This is similar to a filter in
the frequency domain. Here the ``filter'' is the quotient of the desired and
the actual Fourier amplitudes.

In practice, the above null hypothesis is not as interesting as one might like:
Very few of the time series considered for a nonlinear treatment pass even a
simple test for Gaussianity. Therefore we want to consider a more general null
hypothesis including the possibility that the data were measured by an
instantaneous, invertible measurement function $h$ which does not depend on
time $n$. A time series $\{s_n\}, n=1,\ldots,N$ is consistent with this null
hypothesis if there exists an underlying Gaussian linear stochastic signal
$\{x_n\}$ such that $s_n=h(x_n)$ for all $n$. If the null hypothesis is true,
typical realizations of a process which obeys the null are expected to share
the same power spectrum and amplitude distribution. But even within the class
defined by the null hypothesis, different processes will result in different
power spectra and distributions. It is now an essential requirement that the
discrimiating statistics must not mistake these variations for deviations from
the null hypothesis. The tedious way to achieve this is by constructing a
``pivotal'' statistics which is insensitive to these differences. The
alternative we will pursue here is the ``constrained realizations'' approach:
the variations in spectrum and distribution within the class defined by the
null hypothesis are suppressed by constraining the surrogates to have the same
power spectrum as well as the same distribution of values as the data.

In~\cite{surro}, the amplitude adjusted Fourier transform (AAFT) algorithm is
proposed for the testing of this null hypothesis. First, the data $\{s_n\}$ is
rendered Gaussian by rank--ordering according to a set of Gaussian random
numbers. The resulting series $s'_n=g(s_n)$ is Gaussian but follows the
measured time evolution $\{s_n\}$. Now make phase randomized surrogates for
$\{s'_n\}$, call them $\{\tilde{s}'_n\}$. Finally, invert the rescaling $g$ by
rank--ordering $\{\tilde{s}'_n\}$ according to the distribution of the original
data, $\tilde{s}_n=\overline{g}(\tilde{s}'_n)$.

The AAFT algorithm should be correct asymptotically in the limit as
$N\to\infty$.~\cite{rem:AAFT} For finite $N$ however, $\{\tilde{s}_n\}$ and
$\{s_n\}$ have the same distributions of amplitudes by construction, but they
do not usually have the same sample power spectra. One of the reasons is that
the phase randomization procedure performed on $\{s'_n\}$ preserves the
Gaussian distribution only on average.  The fluctuations of $\{\tilde{s}'_n\}$
and $\{s'_n\}$ will differ in detail.  The nonlinearity contained in the
amplitude adjustment procedure ($\overline{g}$ is not equal to $g^{-1}$) will
turn these into a bias in the empirical power spectrum. Such systematic errors
can lead to false rejections of the null hypothesis if a statistic is used
which is sensitive to autocorrelations.  The second reason is that $g$ isn't
really the inverse of the nonlinear measurement function $h$, and instead of
recovering $\{x_n\}$ we will find some other Gaussian series. Even if $\{s_n\}$
were Gaussian, $g$ would not be the identity. Again, the two rescalings will
lead to an altered spectrum.

\begin{figure}
\centerline{\epsffile{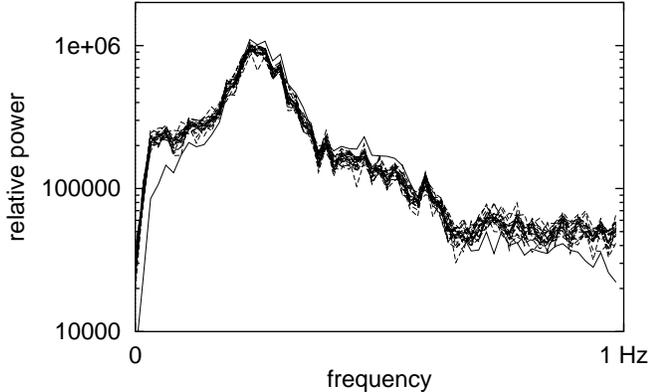}}
\caption[]{Discrepancy of the power spectra of human breath rate data (solid
line) and 19 AAFT surrogates (dashed lines). Here the power spectra have been
   computed with a square window of length~64..\label{fig1}}
\end{figure}

In Fig.~\ref{fig1} we see power spectral estimates of a clinical data set and
of 19 AAFT surrogates.  The data is taken from data set B of the Santa Fe
Institute time series contest~\cite{sfi}. It consists of 4096 samples of the
breath rate of a patient with sleep apnea. The sampling interval is
0.5~seconds. The discrepancy of the spectra is significant. A bias towards a
white spectrum is noted: power is taken away from the main peak to enhance the
low and high frequencies.

The purpose of this letter is to propose an alternative method of producing
surrogate data sets which have the same power spectrum and distribution as a
given data set. We do not expect that these two requirements can be exactly
fulfilled at the same time for finite $N$, except for the trivial solution, a
cyclic shift of the data set itself. We will rather construct sequences which
assume the same values (without replacement) as the data and which have spectra
which are practically indistinguishable from that of the data. We can require a
specific maximal discrepancy in the power spectrum and report a failure if this
accuracy could not be reached.

The algorithm consists of a simple iteration scheme. Store a sorted list of the
values $\{s_n\}$ and the squared amplitudes of the Fourier transform of
$\{s_n\}$, $S^2_k=|\sum_{n=0}^{N-1}s_n e^{i2\pi kn/N}|^2$. Begin with a random
shuffle (without replacement) $\{s^{(0)}_n\}$ of the data.\cite{rem:begin} Now
each iteration consists of two consecutive steps. First $\{s^{(i)}_n\}$ is
brought to the desired sample power spectrum.  This is achieved by taking the
Fourier transform of $\{s^{(i)}_n\}$, replacing the squared amplitudes
$\{S^{2,(i)}_k\}$ by $\{S^2_k\}$ and then transforming back.  The phases of the
complex Fourier components are kept. Thus the first step enforces the correct
spectrum but usually the distribution will be modified. Therefore, as the
second step, rank--order the resulting series in order to assume exactly the
values taken by $\{s_n\}$. Unfortunately, the spectrum of the resulting
$\{s^{(i+1)}_n\}$ will be modified again. Therefore the two steps have to be
repeated several times.

At each iteration stage we can check the remaining discrepancy of the spectrum
and iterate until a given accuracy is reached. For finite $N$ we don't expect
convergence in the strict sense. Eventually, the transformation towards the
correct spectrum will result in a change which is too small to cause a
reordering of the values. Thus after rescaling, the sequence is not changed.

\begin{figure}\centerline{\epsffile{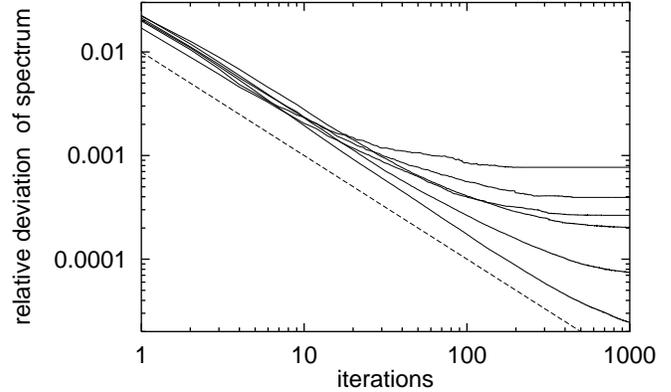}}
\caption[]{Convergence of the iterative scheme to the correct power spectrum
while the distribution is kept fixed. First order AR process with nonlinear
   measurement. The curves were obtained with $N=1024,2048,\ldots,32768$,
   counted from above.  We also show the curve $\propto 1/i$.\label{fig2}}
\end{figure}

\begin{figure}
\centerline{\epsffile{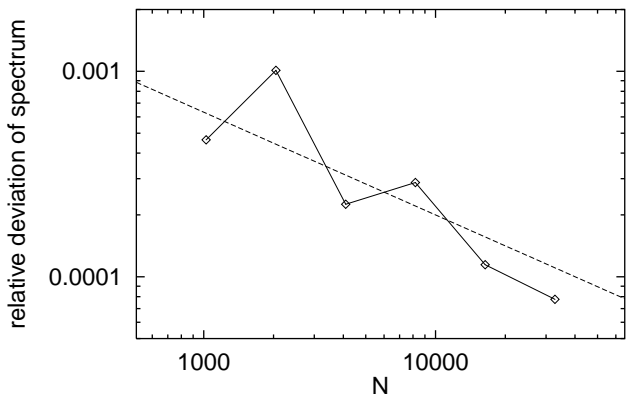}} 
\caption[]{For the same process as used in Fig.~\ref{fig2} we show the
saturation value of the accuracy for the above values of $N$. The straight line
   is $\propto 1/\sqrt{N}$. \label{fig3}}
\end{figure}

In Fig.~\ref{fig2} we show the convergence of the iteration scheme as a
function of the iteration count $i$ and the length of the time series $N$. The
data here was a first order AR process $x_n=0.7x_{n-1}+\eta_n$, measured
through $s_n=x_n^3$. The increments $\eta_n$ are independent Gaussian random
numbers. For each $N=1024,2048,\ldots,32768$ we create a time series and ten
surrogates. In order to quantify the convergence, the spectrum was estimated by
$S^2_k=|\sum_{n=0}^{N-1}s_n e^{i2\pi kn/N}|^2$ and smoothed over 21 frequency
bins, $\hat{S}^2_k=\sum_{j=k-10}^{k+10} S^2_k/21$. Note that for the generation
of surrogates no smoothing is performed. As the (relative) discrepancy of the
spectrum at the $i$--th iteration we use $\sum_{k=0}^{N-1}
(\hat{S}^{(i)}_k-\hat{S}_k)^2/\sum_{k=0}^{N-1} \hat{S}_k^2$. Not surprisingly,
progress is fastest in the first iteration, where the random scramble is
initially brought from its white spectrum to the desired one (the initial
discrepancy of the scramble was $0.2\pm 0.01$ for all cases and is not shown in
Fig.~\ref{fig2}). For $i\ge 1$, the discrepancy of the spectrum decreases
approximately like $1/i$ until an $N$ dependent saturation is reached.  The
saturation value seems to scale like an inverse power of $N$ which depends on
the process. For the data underlying Fig.~\ref{fig2} we find a $1/\sqrt{N}$
dependence, see Fig.~\ref{fig3}. For comparison, the discrepancy for AAFT
surrogates did not fall below 0.015 for all $N$. We have observed similar
scaling behavior for a variety of other linear correlated processes. For data
from a discretized Mackey--Glass equation we found exponential convergence
$\propto \exp(-0.4i)$ before a saturation value was reached which decreases
approximately like $1/N^{3/2}$. Although we found rapid convergence in all
examples we have studied so far, the rate seems to depend both on the
distribution of the data and the nature of the correlations.  The details of
the behavior are not yet understood.

\begin{figure}
\centerline{\epsffile{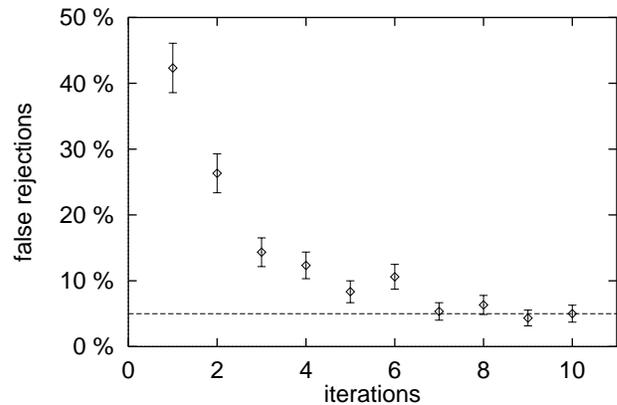}}
\caption[]{Percentage of false rejections as a function of the number of
iterations performed. Horizontal line: nominal rejection rate at the 95\% level
   of significance. In this case, 7 iterations are sufficient to render the
   test accurate. The usual AAFT algorithm yields 66\% false rejections.
   \label{fig4}}
\end{figure}

In order to verify that false rejections are indeed avoided by this scheme we
compared the number of false positives in a test for nonlinearity for the AAFT
algorithm and the iterative scheme, the latter as a function of the number of
iterations. We performed tests on data sets of 2048 points generated by the
instantaneously, monotonously distorted AR process $s_n=x_n\sqrt{|x_n|}$,
$x_n=0.95x_{n-1}+\eta_n$. The discriminating statistic was a nonlinear
prediction error obtained with locally constant fits in two dimensional delay
space. For each test, 19 surrogates were created and the null hypothesis was
rejected at the 95\% level of significance if the prediction error for the data
was smaller then those of the 19 surrogates. The number of false rejections was
estimated by performing 300 independent tests. Instead of the expected 5\%
false positives we found $66\pm 5$\% false rejections with the AAFT
algorithm. Fig.~\ref{fig4} shows the percentage of false rejections as a
function of the number of iterations of the scheme described in this
letter. The correct rejection rate for the 95\% level of significance is
reached after about 7 iterations. This example is particularly dramatic because
of the strong correlations, although the nonlinear rescaling is not very
severe.

Let us make some further remarks on the proposed algorithm. We decided to use
an unwindowed power spectral estimate which puts quite a strong constraint on
the surrogates (the spectrum fixes $N/2$ parameters). Thus it cannot be
excluded that the iterative scheme is able to converge only by also adjusting
the phases of the Fourier transform in a nontrivial way. This might introduce
spurious nonlinearity in the surrogates in which case we can find the confusing
result that there is {\em less} nonlinearity in the data than in the
surrogates. If the null hypothesis is wrong, we expect {\em more} nonlinearity
in the data (better nonlinear predictability, smaller estimated dimension
etc.). Therefore we can always use one--sided tests and thus avoid additional
false rejections. However, spurious structure in the surrogates can diminish
the power of the statistical test.  Since an unwindowed power spectral estimate
shows strong fluctuations within each frequency bin, it seems unnecessary to
require the surrogates to have {\em exactly} the same spectrum as the data,
including the fluctuations.  The variance of the spectral estimate can be
reduced for example by windowing, but the frequency content of the windowing
function introduces an additional bias.

Let us finally remark that although the null hypothesis of a Gaussian linear
process measured by a monotonous function is the most general we have a proper
statistical test for, its rejection does not imply nonlinear dynamics.  For
instance, noninstantaneous measurement functions (e.g., $s_n=x^2_nx_{n-1}$) are
not included and (correctly) lead to a rejection of the null hypothesis,
although the underlying dynamics may be linear. Another example is first
differences of the distorted output from a Gaussian linear process.\cite{dean}

In conclusion, we established an algorithm to provide surrogate data sets
containing random numbers with a given sample power spectrum and a given
distribution of values. The achievable accuracy depends on the nature of the
data and in particular the length of the time series.

We thank James Theiler, Daniel Kaplan, Tim Sauer, Peter Grassberger, and Holger
Kantz for stimulating discussions.  This work was supported by the SFB 237 of
the Deutsche Forschungsgemeinschaft.



\begin{thebibliography}{10}
\bibitem{rem:AAFT} Asymptotically for $N\to\infty$, $g^{-1}$ approaches the
   inverse of the nonlinear measurement function $h$ with mean sqared
   fluctuations of $O(1/N)$. The phase
   randomization preserves the Gaussian distribution in that limit.  Since
   the amplitude adjustments are anticorrelated with the signal, the AAFT
   algorithm introduces a bias towards a flat spectrum which is also expected
   (and numerically found) to be $O(1/N)$.
\bibitem{rem:begin} 
Alternatively but not equivalently, one could start with a usual AAFT
surrogate.
\end{thebibliography}
\end{document}